# Transfer Learning Enabled Transformer based Generative Adversarial Networks (TT-GAN) for Terahertz Channel Modeling and Generating

## Author Information

### Affiliations

**Terahertz Wireless Communications (TWC) Laboratory, Shanghai Jiao Tong University, Shanghai 200240, China.**

Zhengdong Hu, Yuanbo Li, and Chong Han.

### Contributions

Z. Hu, Y. Li, and C. Han conceived the study; Z. Hu and C. Han contributed to the development of transformer based generative adversarial networks and the transfer learning process; Y. Li contributed to the Terahertz channel measurement and implementation of 3GPP method; Z. Hu and C. Han contributed to writing and editing of the manuscript.

### Corresponding author

Correspondence to: Chong Han (ORCID: 0000-0002-9638-4736)

## Abstract

Terahertz (THz) communications, ranging from 100 GHz to 10 THz, are envisioned as a promising technology for 6G and beyond wireless systems. As foundation of designing THz communications, channel modeling and characterization are crucial to scrutinize the potential of the new spectrum. However, current channel modeling and standardization heavily rely on measurements, which are both time-consuming and costly to obtain in the THz band. Here, we propose a Transfer learning enabled Transformer based Generative Adversarial Network (TT-GAN) for THz channel modeling. Specifically, as a fundamental building block, a GAN is exploited to generate channel parameters, which can substitute measurements. To greatly improve the accuracy, the first T, i.e., a transformer structure with a self-attention mechanism is incorporated in GAN. Still incurring errors compared with ground-truth measurement, the second T, i.e., a transfer learning is designed to solve the mismatch between the formulated network and measurement. The proposed TT-GAN can achieve high accuracy in channel modeling, while requiring only rather limited amount of measurement, which is a promising complementary of channel standardization that fundamentally differs from the current techniques that heavily rely on measurement.

## Introduction

With the exponential growth of the number of interconnected devices, the sixth generation (6G) is expected to achieve intelligent connections of everything, anywhere, anytime[1], which demands Tbps wireless data rates. To fulfil the demand, Terahertz (THz) communications gain increasing attention as a vital technology of 6G systems, thanks to the ultra-broad bandwidth ranging from tens of GHz to hundreds of GHz[2-5]. The THz band is promising to address the spectrum scarcity and capacity limitations of current wireless systems, and realize long-awaited applications, extending from metaverse/XR, wireless fronthaul/backhaul, to joint millimeter-level sensing and Tbps-rate communication[6,7].

To design reliable THz wireless systems, an accurate channel model is fundamental to portray the propagation phenomena. However, channel modeling in the THz band is a challenging problem. Due to the high frequencies, new characteristics occur in the THz band, including frequency-selective absorption loss and rough-surface scattering[8], which are not characterized by the existing channel models. Moreover, traditional statistical channel modeling often necessitates a large amount of measurements, which are extremely time-consuming and costly to obtain for THz channel modeling. Therefore, an accurate channel modeling method with very few measurement data for the THz band is pressing but still missing.

With the rising popularity of deep learning (DL), researchers are seeking an answer on how DL methods can be applied in wireless communications, especially in channel modeling[9]. Among different branches of DL methods, the generative adversarial network (GAN)[10-15] has been explored to generate channel parameters that can, at least hopefully, act as a replacement for measurement. This can significantly reduce the challenges and costs of obtaining hundreds and thousands of measurement channels. As an example, a GAN based channel modeling method is proposed and demonstrated over an AWGN channel[13]. Moreover, Xiao et.al.[14] designed a GAN structure to generate channel matrix samples close to the distribution of real channel samples, obtained from clustered delay line (CDL) channel model. Although the motivation and attempts are appreciated, there still exist some limitations of GAN based channel modeling in the current works. Particularly, all of the aforementioned works train the GAN network with large number of simulated channel samples, generated by the conventional channel models. This causes severe mismatch between the channel data generated from the formulated GAN models and the ground-truth measurement data, due to both the inaccuracy of GAN itself and ignorance the features from measurement.

In this paper, we propose a Transfer learning enabled Transformer-based Generative Adversarial Networks (TT-GAN) for THz channel modeling and generating. TT-GAN models the channel by generating spatial-temporal channel parameters, including path gain, phase, delay and angle of the multipaths. With the first T denoting transformer, the transformer structure[16,17] is integrated in T-GAN to improve the accuracy of the basic GAN[18]. The self-attention mechanism introduced in the transformer structure allows T-GAN to focus on important parts of the input channel parameters by giving different attention weights, which can help improve the quality of the representations learned by the model. This can lead to more stable training and better accuracy. Moreover, to tackle the challenge of limited channel measurement in the THz band, the second T that refers to a transfer learning technique[19-22,] is introduced to extract knowledge from scarce THz measurement, which further reduces the discrepancy between the generated channels and measurement.

Overall, design of the proposed TT-GAN is elaborated as follows. To start with, the proposed transformer based GAN (T-GAN) is pre-trained using the simulated dataset, generated by the standard channel model from 3rd generation partnerships project (3GPP)[23]. Furthermore, by transferring the knowledge and fine-tuning the pre-trained T-GAN, the TT-GAN is developed by using the THz measured dataset with only dozens of channels. This can alleviate the demand of large amount of measurement data for training and improve the accuracy of TT-GAN, since the simulated data can serve as a good supplement for the initialization of TT-GAN network. Finally, the proposed TT-GAN can accurately model the channel distribution, generating path loss, delay spread, angular spread and power delay angular profile, which completely portrays the THz channel.

## Results

**THz channel measurement results**

Measurement results are served as ground-truth, for which a channel sounder system supporting high frequencies up to 400 GHz is developed[24]. The system can achieve a time resolution of 66.7 ps, which suggests that it can differentiate between two paths as long as their distance difference is greater than 2 cm.

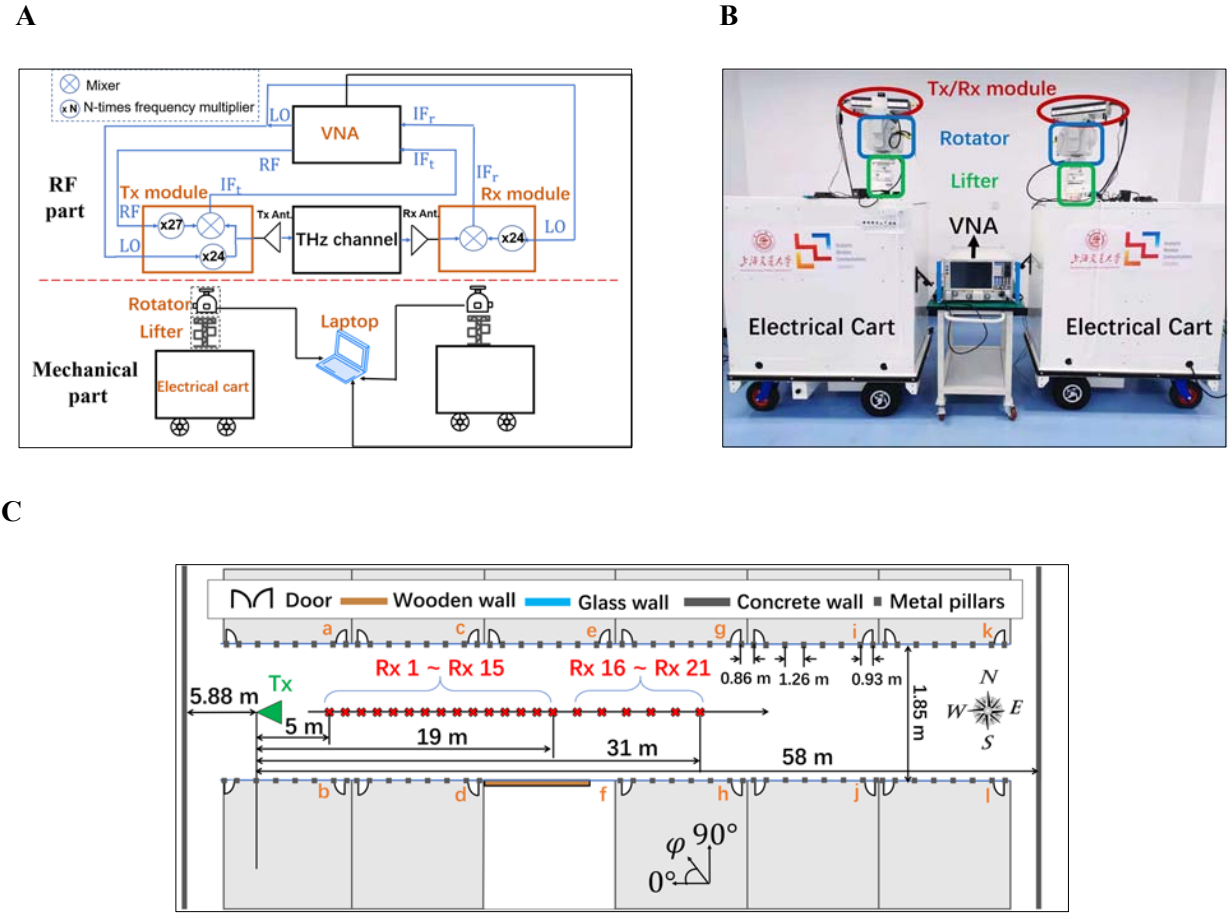

**Fig. 1: Measurement system and layout of measurement campaign.** **A** The block diagram of measurement system with a radio frequency subsystem and a mechanical subsystem. **B** The picture of hardware equipment for the measurement system. **C** The measurement campaign is carried out in the corridor, with the glass walls connected by metal pillars on the both sides. The materials of glass and metal can significantly impact the propagation of THz signals, affecting reflection and scattering. This setup ensures our study's relevance to practical indoor THz communication scenarios.

The structure of the measurement system is shown in Fig. 1**A**, which consists of a radio frequency (RF) subsystem and a mechanical subsystem. The RF component manages the transmission of the RF signal to measure the THz channels, while the mechanical component facilitates the adjustment of positions, heights, and steering angles of the transceivers. The hardware implementation of the measurement system is shown in Fig. 1**B**. To calculate the channel transfer function (CTF) of the THz channel, the Vector Network Analyzer (VNA) initially generates an RF signal with frequencies ranging from 9.6926 GHz to 14.8418 GHz, which is then directed to the transmitter (Tx) module. In the Tx module, the RF signal passes through a 27-times frequency multiplier to produce a THz signal with frequencies within 260–400 GHz. Simultaneously, a 260.279–400.279 GHz local oscillation (LO) signal is generated by performing a 24-times frequency multiplication on a 10.8450–16.6783 GHz signal. By combining the THz RF signal and the LO signal, a 279 MHz intermediate frequency (IF) signal IFt is obtained, which is then sent back to the VNA. The THz RF signal, on the other hand, travels through the Tx antenna, across the THz channel, and finally arrives at the receiver end. Here, it mixes with an LO signal to transmit a 279 MHz IF signal IFr back to the VNA. Ultimately, the ratio between the frequency responses of IFr and IFt determines the channel transfer function of the THz channel.

Based on the developed measurement system, the measurement campaign is conducted in an indoor corridor scenario at 306-321 GHz, as depicted in Fig. 1**C**. The location of transmitter is fixed at the left end of the corridor, while 21 receiver positions are located along the corridor. To receive THz channel from different directions, the receiver scans the spatial domain at a resolution of 10 degrees, with azimuth planes of 0° to 360° and elevation planes of −20° to 20°. The measurement dataset consists of 21 channels, each represented as a superposition of multi-path components (MPCs). The measured channel data exhibits clear sparsity, with only 6 to 8 main paths being significant. The average delay spread is 10.94 ns, and the average angular spread is 30.99°. Moreover, the average path loss exponent (PLE) is 1.5138, significantly lower than the free-space PLE of 2. This reduction is attributed to the waveguide effect in the corridor.

**Transfer learning enabled transformer-based generative adversarial networks**

We firstly formulate the channel modeling as a channel parameter generating problem. The THz channel can be represented as

$$h(\tau) = \sum_{l=0}^{L-1} \alpha_l e^{j\phi_l} \delta(\tau - \tau_l), \quad (1)$$

where $l = 1, \cdots, L$ indexes the multi-path components (MPCs), $\alpha_l$ denotes the path gain of the $l^{th}$ MPC, $\phi_l$ represents the random phase, and $\tau_l$ denotes the delay of the $l^{th}$ MPC. Every MPC can be characterized by a set of parameters as

$$\mathbf{x}_l = [\alpha_l, \tau_l, \theta_l, \psi_l], \quad (2)$$

where $\mathbf{x}_l$ denotes $l^{th}$ MPC, the $\theta_l$ and $\psi_l$ represent the azimuth angle of arrival (AoA) and elevation angle of arrival (EOA), respectively. Then, the THz channel can be characterized by

$$\mathbf{x} = [\mathbf{x}_1, \mathbf{x}_2, \cdots, \mathbf{x}_L], \quad (3)$$

where the number of MPCs $L$ is set as 15, considering the sparsity of THz channel[1]. The problem of channel modeling can then be described as the generation of channel parameters that forms a distribution of channels. The generating process can be represented by the function

$$\hat{\mathbf{x}} = G(\mathbf{z}|c), \quad (4)$$

where $\mathbf{z}$ denotes a random vector sampled from a normal distribution, the variable $c$ is the condition information representing the distance between the transmitter and receiver. Through the function G, the target channel distribution $p_r(\mathbf{x}|c)$ conditioned on the distance can be approximated by the generated distribution $p_g(\hat{\mathbf{x}}|c)$.

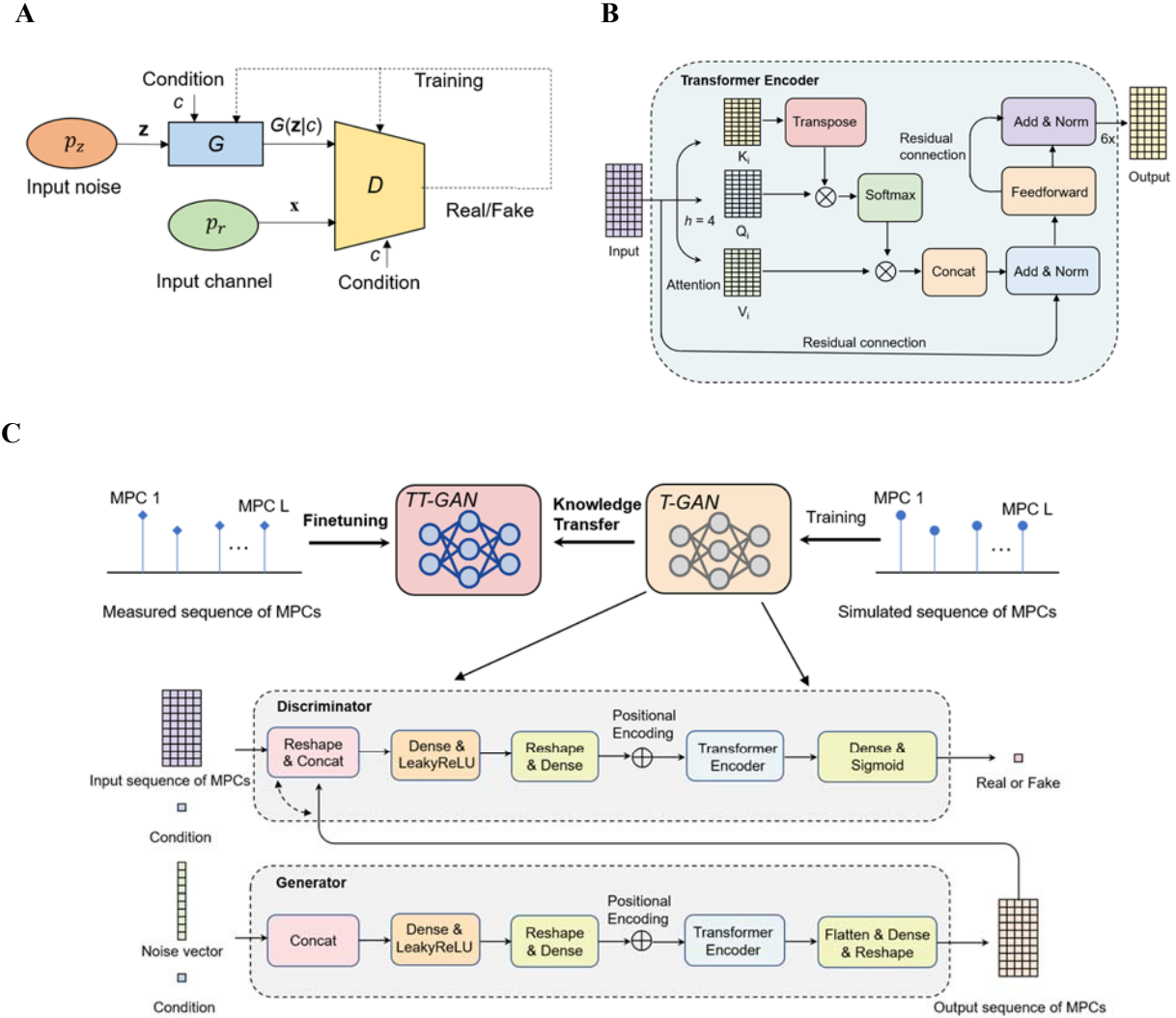

**Fig. 2**: **Framework of T-GAN and TT-GAN. A** In the framework of T-GAN, the generator *G* aims to generate fake channel *G*(**z**|*c*) with the input of noise vector **z** and distance information *c*, while the discriminator *D* tries to distinguish the input real channel **x** or fake channel *G*(**z**|*c*) as real or fake. The $p_r$ and $p_z$ represent the distributions of real channels and noise vector, respectively. **B** The transformer encoder structure incorporated in T-GAN. **C** In the framework of TT-GAN, the T-GAN is firstly trained on the simulated channels, represented as the sequence of MPCs. Then, the knowledge of T-GAN is transferred

to TT-GAN, and the TT-GAN is further finetuned by measured channels. The network structure of T-GAN is shown in the below of the figure.

**T-GAN.** The T-GAN network can be exploited to generate channel parameters that forms a distribution of channels. The framework of the proposed T-GAN is shown in Fig. 2**A**, which consists of two sub-networks, namely, generator $G$ and discriminator $D$. The generator is aimed at generating the fake channel $G(\mathbf{z}|c)$ conditioned on the distance information $c$ to fool the discriminator, while the discriminator serves as a classifier, trying to distinguish between the real channel $\mathbf{x}$ and fake channel $G(\mathbf{z}|c)$. The two networks are then trained in an adversarial manner, which can be considered as a two-player zero-sum minimax game. Specifically, the training objective can be represented by

$$\min_G \max_D \mathbb{E}_{\mathbf{x} \sim p_r}[\log D(\mathbf{x}|c)] + \mathbb{E}_{\mathbf{z} \sim p_z}\left[\log\left(1 - D(G(\mathbf{z}|c))\right)\right], \quad (5)$$

where $p_r$ and $p_z$ represent the distributions of real channels and noise vector, respectively. The generator minimizes $(1 - D(G(\mathbf{z}|c)))$ that represents the probability of the generated channel detected as fake, while the discriminator maximizes this probability. Therefore, the generator and discriminator compete against each other with the opposite objectives in the training process. Through the adversarial training, the Nash equilibrium can be achieved, such that the generator and discriminator cannot improve their objectives by changing only their own network. However, training with the objective function in equation (5) is unstable, since the training objective is potentially not continuous with respect to the generator's parameters[18]. Therefore, the improved version of GAN, namely, Wasserstein GAN[18] with gradient penalty is adopted. The modified objective function is expressed as

$$\min_G \max_D \mathbb{E}_{\mathbf{x} \sim p_r}[\log D(\mathbf{x}|c)] + \mathbb{E}_{\mathbf{z} \sim p_z}\left[\log\left(1 - D(G(\mathbf{z}|c))\right)\right] + \lambda \mathbb{E}_{\tilde{\mathbf{x}}}[(\|\nabla_{\tilde{\mathbf{x}}} D(\tilde{\mathbf{x}}|c)\| - 1)^2], \quad (6)$$

where the last term is the gradient penalty term to enforce Lipschitz constraint that the gradient of the network is upper-bounded by a maximum value, the symbol $\tilde{\mathbf{x}}$ is the uniformly sampled point between the points of x and $G(\mathbf{z}|c)$. Moreover, the parameter $\lambda$ is the penalty coefficient.

In T-GAN, the channel is input as a sequence of MPCs as in equation (3). Hence, the transformer encoder structure can be utilized to capture the dependencies among the sequence of MPCs by the self-attention mechanism. Self-attention mechanism allows the T-GAN model to flexibly utilize the most relevant parts of the input MPC sequence, by a weighted combination of all the encoded MPC vectors. Through the self-attention mechanism, T-GAN can relate the different positions of a single sequence to calculate its representation with improved quality. This can lead to more stable training and better accuracy performance. The detailed explanation of the transformer encoder structure and the full network structure of T-GAN are given in "Methods".

**Transfer Learning.** The framework of TT-GAN is shown in Fig. 2**C**. The T-GAN is firstly pre-trained by feeding the simulated dataset generated by QuaDRiGa[25], with the extracted statistics from THz measurement. Specifically, QuaDRiGa is an open-sourced implementation of 3GPP TR 38.901 model. The extracted statistics from THz measurement include path loss exponent, the mean and standard deviation of K-factor, the delay spread, and angular spread as well as the correlation matrix. However, the simulated dataset cannot match the measurement accurately, which causes mismatch between the T-GAN model and ground-truth measurement.

To tackle this mismatch problem, the transfer learning is exploited to transfer the knowledge from T-GAN to TT-GAN. Specifically, the TT-GAN is initialized with the weights of the T-GAN pre-trained on the simulated channels. It is worth noting that the generator and discriminator in the T-GAN are both transferred to TT-GAN, which can yield the better performance in generating high quality samples and fast convergence, compared with transferring only the generator or the discriminator[22].

The TT-GAN model is then fine-tuned[22] using the measurement dataset, with only a small amount of data. It involves taking a pre-trained model and updating its parameters using a new dataset that is specific to the task at hand. By adjusting the weights of the pre-trained model, the fine-tuning process allows the new model to learn task-specific features while retaining the knowledge gained from the original training.

This approach is particularly useful when the new dataset is small or similar to the original dataset, as it can save time and computational resources compared to training a new model from scratch.

During the fine-tuning process, regularization is an important technique to avoid the over-fitting problem when training on the small dataset. Moreover, to preserve the knowledge learned in the initial model, the L2-SP regularization is applied to enforce the fine-tuned model close to the initial model. The L2-SP regularization can be represented as

$$\Omega(w) = \frac{\alpha}{2}\|w - w^0\|_2, \tag{7}$$

Where $w$ and $w^0$ denote the parameters of the fine-tuned network and the initial network respectively. Moreover, $\alpha$ is the regularization parameter. By using the L2-SP regularization term, the search space of the fine-tuned network is constrained around the start point (SP) of initial network, which helps keep the acquired knowledge in the initial model. As a result, through this fine-tuning process, transfer learning enables TT-GAN to effectively learn the channel distribution from measurement

**Performance evaluation of TT-GAN**

To evaluate the performance of TT-GAN, a comprehensive evaluation is conducted based on several key metrics essential for channel modeling. The metrics used include delay spread, angular spread, path loss, and power delay angular spread. These characteristics of the generated channels are compared with those of measured channels to validate the model's accuracy. The TT-GAN generated channels have an average delay spread of 14.37 ns, an average angular spread of 33.84°, and a path loss exponent of 1.4908, closely matching the measured values of 10.94 ns, 30.99°, and 1.5138, respectively. Additionally, TT-GAN is benchmarked against several channel modeling methods, namely ray-tracing, a basic GAN, and the previously mentioned T-GAN without transfer learning. These comparisons are conducted to evaluate the accuracy of each method against ground-truth measurement. Moreover, the impact of the measurement dataset size on the accuracy of TT-GAN is analyzed. This analysis aims to understand the robustness of the model when trained with little measurement data. Lastly, the computational complexities of all the aforementioned methods are compared. This comparison highlights the computational efficiency of TT-GAN for channel modeling.

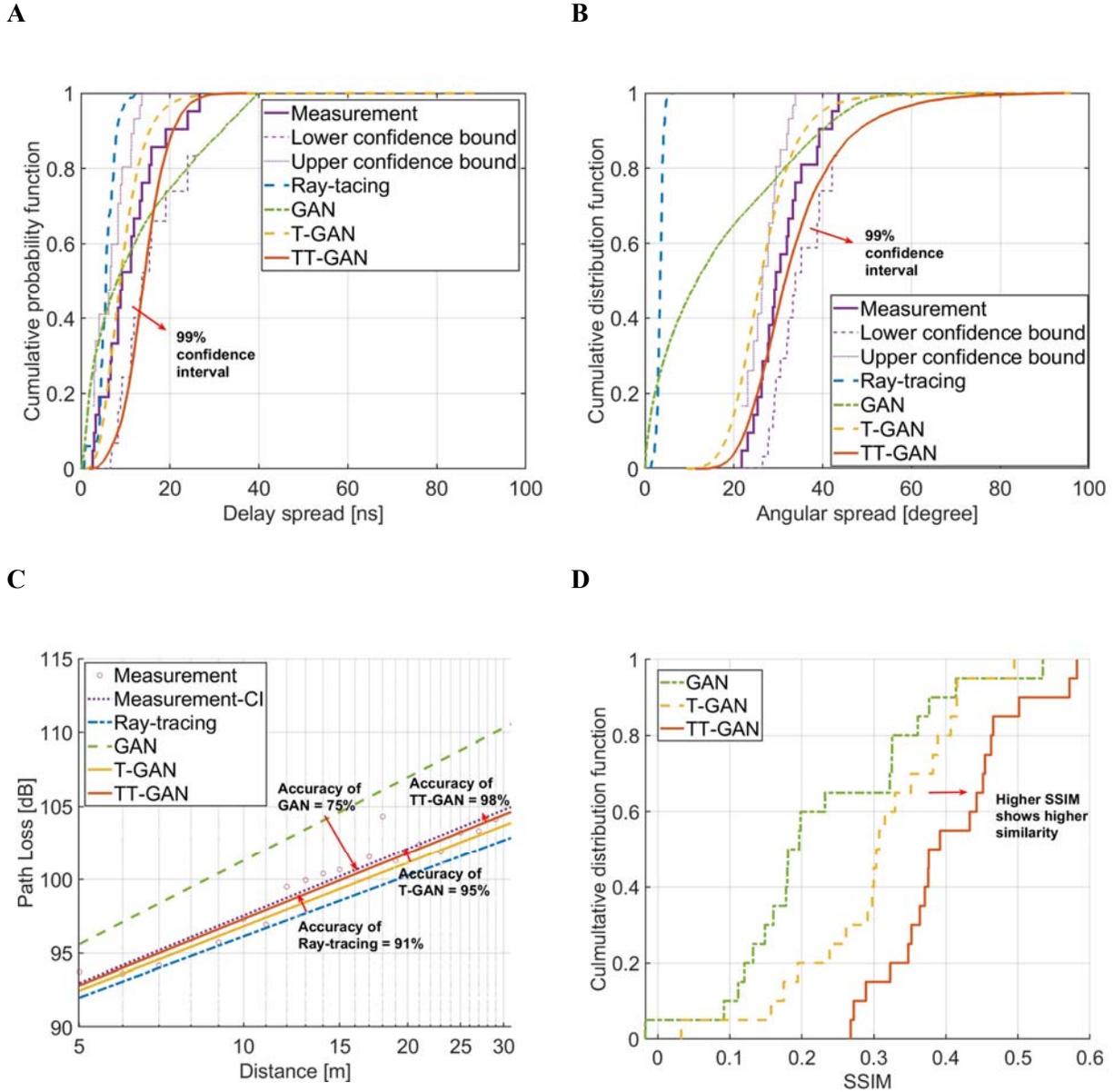

**Fig. 3: Performance evaluation of TT-GAN. A** The cumulative distribution function (CDF) plots of delay spread for T-GAN and TT-GAN lie within the 99% confidence interval of the measurement CDF, which is the interval between the lower confidence bound and upper confidence bound. This shows that the T-GAN and TT-GAN can accurately model the delay spread of measurement channels. **B** The CDFs of angular

spread for T-GAN and TT-GAN fall within the 99% confidence interval of the measurement CDF, which indicates that the T-GAN and TT-GAN can accurately model the angular spread of measurement channels. **C** The path loss of measured and generated channels is fitted with the close-in free space reference distance (CI) model. The model fitted with TT-GAN channels achieves the highest accuracy of 98% in predicting the measured path loss exponent. This shows that TT-GAN outperforms other methods in predicting the path loss exponent. **D** The Structural Similarity Index Measure (SSIM) assesses the similarity between generated and measured power delay angular profiles (PDAPs). A higher SSIM value indicates greater similarity between the PDAPs. TT-GAN achieves higher SSIM values compared to GAN and T-GAN methods, demonstrating that TT-GAN can effectively learn from measurement to enhance similarity.

**Delay spread.** The delay spread characterizes the power dispersion of multi-path components in the temporal domain. It is an important metric to measure the small-scale fading, which can be computed by

$$\bar{\tau} = \frac{\sum_{i=0}^{N_\tau} i\Delta\tau P_\tau(i)}{\sum_{i=0}^{N_\tau} P_\tau(i)}, \tag{8-1}$$

$$\tau_{rms} = \sqrt{\frac{\sum_{i=0}^{N_\tau}(i\Delta\tau - \bar{\tau})^2 P_\tau(i)}{\sum_{i=0}^{N_\tau} P_\tau(i)}}, \tag{8-2}$$

where $N_\tau$ denotes the number of sampling points in the temporal domain, $\bar{\tau}$ denotes the mean delay weighted by the power, $\tau_{rms}$ refers to the root-mean-square (RMS) delay spread, $\Delta\tau$ denotes the sampling time interval, and $P_\tau(i)$ denotes the power at the delay of $i\Delta\tau$. The cumulative distribution function (CDF) plot of delay spread for the original and generated channels is depicted in Fig. 3**A**. It can be observed that there exists clear deviation between the ray-tracing result and the measurement. The ray-tracing method models the propagation of electromagnetic waves based on the approximation of the Maxwell equations and geometric optics, which requires detailed environment geometry information. In the measured corridor scenario, there are lots of metal pillars on the both sides, incurring strong scattering effects, which cannot be captured accurately by ray-tracing method. This leads to the poor performance of ray-tracing. Moreover, the CDF of delay spread for GAN trained directly on the measured dataset shows significant differences from the measurement. This is because in the case of a small measurement dataset with 21 channels, the training of GAN is unstable, resulting in GAN unable to learn the property of delay spread well. By comparison, the CDFs of delay spread for channels generated by T-GAN and TT-GAN closely match the measurement and fall within the 99% confidence interval of the measurement CDF, with T-GAN modeling the delay spread slightly better than TT-GAN. On one hand, T-GAN is trained on a simulated dataset that uses extracted statistics of delay spread from measurement. This enables T-GAN to effectively learn the delay spread from a large number of simulated channels. On the other hand, TT-GAN is trained with a small amount of measurement data. To mitigate overfitting problem, L2-SP regularization is applied, enforcing the fine-tuned TT-GAN model to remain close to the initial T-GAN model. This regularization acts as a constraint for the optimization problem of channel modeling. While TT-GAN improves in modeling path loss exponent and power delay angular profile through network training as discussed later, it may not achieve joint improvement of the delay spread and might even degrade slightly due to the regularization constraint.

**Angular spread**. The angular spread describes how the power scatters in the spatial domain, which can be represented by

$$\bar{\theta} = \frac{\sum_{i=0}^{N_\theta} i\Delta\theta P_\theta(i)}{\sum_{i=0}^{N_\theta} P_\theta(i)}, \qquad (9-1)$$

$$\theta_{rms} = \sqrt{\frac{\sum_{i=0}^{N_\theta}(i\Delta\theta - \bar{\theta})^2 P_\theta(i)}{\sum_{i=0}^{N_\theta} P_\theta(i)}}, \qquad (9-2)$$

where $N_\theta$ denotes the number of sampling points in the spatial domain, $\bar{\theta}$ denotes the angle weighted by the power, $\theta_{rms}$ refers to the RMS angular spread, $\Delta\theta$ defines the angle interval, and $P_\theta(i)$ refers to the power at the AoA of $i\Delta\theta$. The CDF plot of angular spread for the original and generated channels is depicted in Fig. 3**B**. There is a noticeable deviation between the ray-tracing results and the measurement. This occurs because the ray-tracing method cannot accurately capture the intricate scattering effects caused by the numerous metal pillars in the measured corridor scenario. The CDFs of angular spread for the generated channels for T-GAN and TT-GAN have a good agreement with the measured channels, falling within the 99% confidence interval of the measured CDF. This suggests that T-GAN and the proposed TT-GAN can well capture the statistics of angular spread in the spatial domain. As stated before, T-GAN and TT-GAN can achieve a good performance in angular spread thanks to the powerful learning ability of the designed transformer based GAN network.

**Path loss.** The path loss is the reduction in power of electromagnetic wave after transmission, which can be calculated by dividing the transmitted power by the received power. To characterize the path loss, a close-in free space reference distance (CI) model is developed, which can be represented by

$$\text{PL}_{\text{CI}}[\text{dB}] = 10 \times \text{PLE} \times \log_{10}\left(\frac{d}{d_0}\right) + \text{FSPL}(d_0), \tag{10}$$

where PLE is the path loss exponent, $d$ represents the Euclidean distance between transmitter (Tx) and receiver (Rx), $d_0$ denotes the reference distance which is selected as 1 m in this work. Moreover, the free-space path loss (FSPL) is calculated by invoking the Friis' law, given by,

$$\text{FSPL}(d_0, f) = -20 \log_{10}\left(\frac{c}{4\pi f d_0}\right), \tag{11}$$

where $c$ denotes the speed of light, $f$ represents the frequency. Then, the CI model is fitted with the measurement channels and the generated channels, respectively, by minimizing the least square error. As can be observed in Fig. 3**C**, the CI model effectively characterizes the relationship between path loss and TX/Rx separation distance. Specifically, the PLEs equal to 1.5138, 1.3725, 1.9331, 1.4408, 1.4908 for measurement, ray-tracing, GAN, T-GAN and TT-GAN, respectively. Among various modeling methods, the GAN exhibits the poorest performance with accuracy of 75%, due to the unstable training problem with a small measurement dataset. Moreover, the PLE of raytracing shows a clear derivation from the measurement with the accuracy of 91%, since the spatial information of the environment and the material properties cannot be obtained precisely. By comparison, T-GAN and TT-GAN learn the PLEs from the input training channels, and the PLE results are very close to the measured PLE with the accuracy of 95% and 98%, respectively. Specifically, T-GAN is trained based on the generated channels by 3GPP, and can well extract the knowledge of PLE from 3GPP based on the powerful transformer structure. Afterwards, TT-GAN inherits the knowledge from T-GAN, and further improves the performance of T-GAN by fine-tuning with calibration from the 21 measured channels. The accuracy of TT-GAN in predicting the PLE increases from 95% to 98% compared with T-GAN, which shows the performance gain brought by the transfer learning technique with measurement.

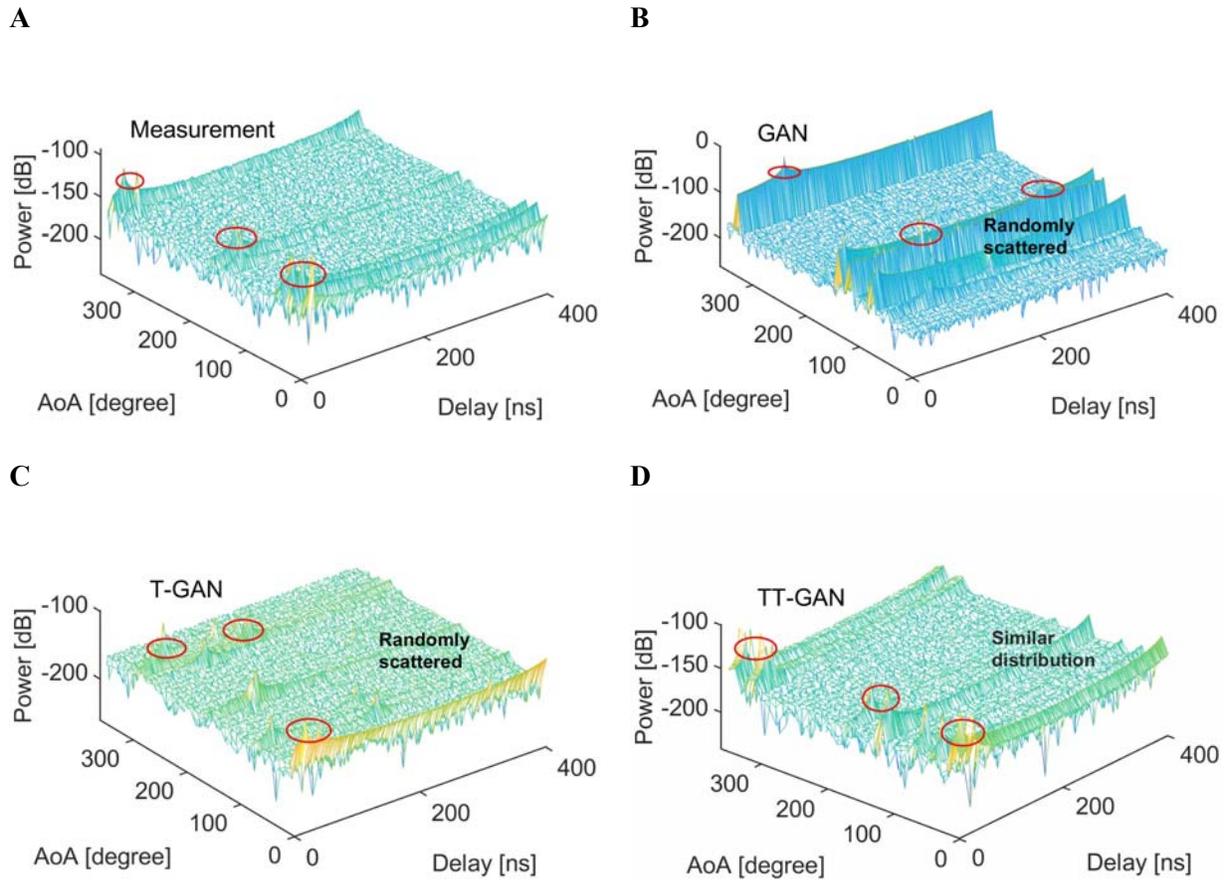

**Fig. 4: Power delay angular profile for measurement, GAN, T-GAN and TT-GAN. A** The MPCs of measured PDAP circled by red ellipses, are mainly distributed around the AoA angles of 0 degree, 180 degree and 360 degree. **B** The range of power for GAN generated PDAP is from -200 dB to 0 dB, which is different from the range of -200 dB to -100 dB for measured PDAP. This indicates a substantial disparity

between the measured and GAN-generated PDAPs. **C** The MPCs in the PDAP generated by T-GAN are randomly dispersed, contrasting with the distribution of MPCs in the measured PDAP. This discrepancy highlights a mismatch between the measurement and the T-GAN. **D** The MPCs in the PDAP generated by TT-GAN have a similar distribution to those in the measured PDAP, clustering around angles of 0 degree, 180 degree, and 360 degree. This demonstrates TT-GAN's ability in accurately modeling the measured channels.

**Power delay angular profile.** The PDAP characterizes the distribution of power in the spatial-temporal domain. In the experiment, the PDAPs for the measured and generated channels are compared as in Fig. 4. The red ellipses circle the region of peak power in the PDAP, which corresponds to the MPCs. It can be observed that PDAP generated by GAN differs significantly from the measured PDAP in terms of power range and MPC distribution. This shows the infeasibility of directly training GAN with the limited measurement dataset. Moreover, the MPCs in PDAP generated by T-GAN are dispersed randomly in the spatial-temporal domain, compared with measurement. The reason is that the T-GAN is based on the 3GPP simulated dataset, which considers only the separation between the Tx and Rx, without relying on prior knowledge of the propagation environment's geometry. Each MPC in 3GPP PDAP is assigned randomly the generated time-of-arrival and angle-of-arrival values based on pre-defined parameter statistics. To stand out, the PDAP generated by TT-GAN has a similar distribution of MPCs to the measured PDAP. This is attributed to TT-GAN's ability to utilize transfer learning technique to learn spatial information from measured channels, thereby improving channel modeling performance.

Moreover, to measure the similarity of PDAP quantitatively, Structure Similarity Index Measure (SSIM) is introduced, which is widely applied to evaluate the quality and similarity of images[26]. The range of SSIM is from 0 to 1, and the value of SSIM is larger when the similarity between images is higher. The PDAPs of the generated channels are compared with the measured channels at the same distance. The CDF of SSIM is shown in Fig. 3**D**. The average SSIMs of GAN, T-GAN, TT-GAN are 0.2286, 0.3029 and 0.4047, respectively. The proposed TT-GAN can achieve a higher SSIM value than GAN and T-GAN methods. This further demonstrates the good performance of TT-GAN in modeling the channels, which outperforms the GAN and T-GAN in terms of the PDAP. By comparison, the T-GAN based on 3GPP cannot achieve as good performance as TT-GAN due to the mismatch between 3GPP and measurement, which shows the necessity of utilizing measurement into the training of GAN with transfer learning technique.

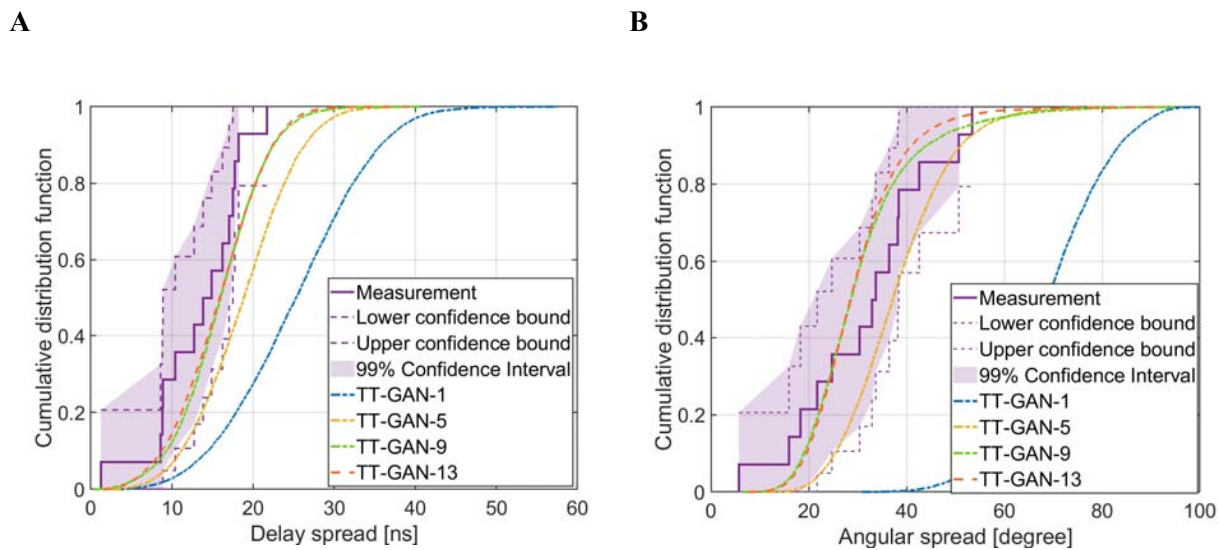

**Fig. 5: Delay spread and angular spread of TT-GAN with varying measurement dataset sizes. A** The CDFs of delay spread for TT-GAN gradually deviate from the CDF of measurement, with the decreasing of dataset sizes. When the dataset size is lower than 9, the CDFs for TT-GAN fall outside the 99% confidence interval of the measurement CDF. This shows that TT-GAN cannot accurately model the angular spread with less than 9 channel samples. **B** The CDFs of angular spread for TT-GAN diverge from the measurement CDF as the dataset size decreases to less than 5. Conversely, this shows that TT-GAN can accurately model the angular spread with a minimum of 5 channel samples.

**Dataset size.** When transitioning to a new environment with limited data, the pre-trained T-GAN model struggles to adapt to different channel distributions. Instead, TT-GAN offers flexibility and efficiency in handling diverse environments by fine-tuning on scarce measurement. To evaluate how the size of the measurement dataset affects TT-GAN's performance, the metrics of delay spread and angular spread are assessed with varying measurement dataset sizes in a new indoor environment[27]. The CDFs of delay spread and angular spread with the decreasing data sizes are shown in Fig. **5A** and Fig. **5B**, respectively. In Fig. **5A**, the CDFs of delay spread for TT-GAN progressively diverge from the measurement CDF as the dataset size decreases. Once the dataset size decreases below 9, TT-GAN's CDF falls outside the 99% confidence interval of the measurement CDF. In Fig. **5B**, the CDFs of angular spread gradually deviate from the measurement as the dataset size decreases to less than 5 samples. Overall, the experiments demonstrate that a minimum of 9 data points from a new environment to be incorporated into the transfer learning process could guarantee high accuracy of the TT-GAN channel model, as evidenced by the CDFs of both delay spread and angular spread for TT-GAN falling within the 99% confidence intervals of the measured CDFs.

**Computation complexity**. Since the ray-tracing and 3GPP methods are implemented using the software programs of Wireless Insite[28] and QuaDRiGa respectively, their computation complexities are hard to be mathematically represented. Therefore, the computation complexities of the aforementioned methods are evaluated by measuring the real computation time required to generate 10000 channels. The basic GAN without the transformer structure has the shortest computation time of 0.44 seconds, while the accuracy of GAN is clearly inferior to T-GAN and TT-GAN as mentioned before. By contrast, the computation time for T-GAN and TT-GAN sharing the same network structure is 1.15 seconds, which is still relatively fast. By comparison, the ray-tracing and 3GPP methods are computationally intensive, which take 1423.58 and 263.82 seconds, respectively. Therefore, TT-GAN demonstrates a clear advantage over the traditional ray-tracing and 3GPP methods in terms of computation speed.

## Discussion

In this paper, we proposed a TT-GAN based THz channel modeling method, which exploits the advantage of GAN in modeling the complex distribution. To improve the performance of GAN networks, T-GAN is firstly developed which integrates the transformer structure. By leveraging the self-attention mechanism in the transformer structure, the T-GAN can effectively identify and focus on crucial aspects of the input channel parameters. This results in high-quality representations learned by the model, leading to more stable training and superior performance. Moreover, transfer learning is deployed by transferring the knowledge from a source task to improve generalization about the target task with limited measurement data. By transferring the knowledge and fine-tuning the pre-trained T-GAN, the TT-GAN is developed by using the THz measured dataset with a small amount.

Then, we evaluate the performance of TT-GAN with the THz measurement data as ground truth. The results show that TT-GAN can achieve high accuracy in channel modeling with rather limited channel data. Specifically, the proposed TT-GAN, can accurately capture the delay spread, angular spread and path loss of the THz channel. Moreover, we compare the simulated PDAP to the measured PDAP in terms of SSIM. A good value of SSIM is achieved when the power, delay, and angle of the simulated paths are consistent with those of the measured paths. Numerical results demonstrate TT-GAN outperforms other methods in terms of SSIM.

With its channel generating capabilities, TT-GAN can produce channel samples from scarce THz measurement while maintaining accurate channel statistics. Using channels generated by TT-GAN, more accurate models can be constructed, enhancing overall communication system efficiency. Additionally, as DL methods become more popular in wireless communication, TT-GAN can provide large quantities of data that closely resemble real measurement, enhancing DL-based systems. For example, TT-GAN can simulate channels, bridging the gap between transmitter and receiver in DL-based end-to-end systems.

## Methods

**Transformer encoder structure**

As depicted in Fig. 2**B**, the transformer encoder consists of 6 stacked identical layers. Every identical layer can be further divided into two sub-layers, multi-head attention layer and feed-forward layer. In both of the two sub-layers, the residual connection is applied by adding the input and the output of the sub-layer represented by x + Sublayer(x). Moreover, the two sub-layers are followed by layer normalization, which can normalize the input and improve the stability of training.

In the multi-head attention layer, multiple attention layers are applied to the input channel in parallel, so that the model can capture the information of the channel in different subspaces. The implementation of a single attention layer is introduced first. Considering an input channel $\mathbf{X} = (\mathbf{x}_1, \cdots, \mathbf{x}_L) \in \mathbb{R}^{L \times d_x}$, it is composed of $L$ MPCs and every MPC is represented by a vector $\mathbf{x}_l \in \mathbb{R}^{1 \times d_x}$. Firstly, every MPC in the sequence is transformed by

$$\mathbf{q}_l = \mathbf{x}_l \mathbf{W}^q, \qquad (12-1)$$

$$\mathbf{k}_l = \mathbf{x}_l \mathbf{W}^k, \qquad (12-2)$$

$$\mathbf{v}_l = \mathbf{x}_l \mathbf{W}^v, \qquad (12-3)$$

where $\mathbf{W}^q \in \mathbb{R}^{d_x \times d_k}$, $\mathbf{W}^k \in \mathbb{R}^{d_x \times d_k}$, $\mathbf{W}^v \in \mathbb{R}^{d_x \times d_v}$ are the learned transformation parameters. The symbols $\mathbf{q}_l \in \mathbb{R}^{1 \times d_k}$, $\mathbf{k}_l \in \mathbb{R}^{1 \times d_k}$ and $\mathbf{v}_l \in \mathbb{R}^{1 \times d_v}$ denote query, key and value respectively. The correlation between the query vector and the key vector shows how much attention should be paid to the value vector in the output. To give a concise representation, the vectors are packed into matrices represented by

$$\mathbf{Q} = \mathbf{X}\mathbf{W}^q, \qquad (13-1)$$

$$\mathbf{K} = \mathbf{X}\mathbf{W}^k, \qquad (13-2)$$

$$\mathbf{V} = \mathbf{X}\mathbf{W}^v, \qquad (13-3)$$

where $\mathbf{Q} \in \mathbb{R}^{L \times d_k}$, $\mathbf{K} \in \mathbb{R}^{L \times d_k}$ and $\mathbf{V} \in \mathbb{R}^{L \times d_v}$ are the matrix representations of query, key and value. Then, the output can be calculated as

$$\text{Attention}(\mathbf{Q}, \mathbf{K}, \mathbf{V}) = \text{softmax}\left(\frac{\mathbf{Q}\mathbf{K}^T}{\sqrt{d_k}}\right)\mathbf{V}, \tag{14}$$

where $\text{Attention}(\mathbf{Q}, \mathbf{K}, \mathbf{V}) \in \mathbb{R}^{L \times d_v}$ is the output of the attention layer, the term $\text{softmax}\left(\frac{\mathbf{Q}\mathbf{K}^T}{\sqrt{d_k}}\right)$ is the calculated attention matrix assigned to the value vector in matrix $\mathbf{V}$. The softmax is the operation for normalizing the attention weights, defined as

$$\text{softmax}(x) = \frac{e^{x_i}}{\sum e^{x_i}}, \tag{15}$$

where $x_i$ is the element in vector $\mathbf{x}$, and the softmax operation ensures that the sum of the output equals one. With the single attention layer introduced, the multi-head attention layer is formed by concatenating the result of $h = 4$ attention layers, which can be represented by

$$\text{Head}_i = \text{Attention}(\mathbf{Q}_i, \mathbf{K}_i, \mathbf{V}_i), \tag{16-1}$$

$$\mathbf{X}^o = \text{Concat}(\text{Head}_1, \text{Head}_2, \cdots, \text{Head}_h), \tag{16-2}$$

where $i = 1, \cdots, 4$ indexes the attention layer, the term $\text{Head}_i \in \mathbb{R}^{L \times d_v}$ denotes the result of the $i^{\text{th}}$ parallel attention layer, $\mathbf{W}^o \in \mathbb{R}^{hd_v \times d_x}$ is the linear matrix that transforms the concatenated result $\mathbb{R}^{L \times hd_v}$ into the output $\mathbf{X}^o \in \mathbb{R}^{L \times d_x}$.

The output of the multi-head attention layer is then passed to the feedforward layer, which is just two dense layers with ReLU activation. The ReLU activation function is defined as

$$f(x) = \max(0, x). \tag{17}$$

Then, the feedforward operation can be characterized by

$$\text{FFN}(\mathbf{X}^o) = \max(0, \mathbf{X}^o \mathbf{W}_1 + \mathbf{b}_1)\mathbf{W}_2 + \mathbf{b}_2, \tag{18}$$

where $\mathbf{X}^0 \in \mathbb{R}^{L \times d_x}$ denotes the input to the feedforward layer. Moreover, $\mathbf{W}_1 \in \mathbb{R}^{d_x \times d_x}$ and $\mathbf{W}_2 \in \mathbb{R}^{d_x \times d_x}$ are the linear transformation matrices, and $\mathbf{b}_1 \in \mathbb{R}^{d_x \times 1}$ and $\mathbf{b}_2 \in \mathbb{R}^{d_x \times 1}$ are the bias terms for the two dense layers.

**Network structure of T-GAN**

The structure of the proposed T-GAN is shown in the bottom part of Fig. 2C, which consists of two sub-networks, namely, generator $G$ and discriminator $D$. The input to the generator includes the noise vector $\mathbf{z} \in \mathbb{R}^{32 \times 1}$ and the condition variable $c \in \mathbb{R}^{1 \times 1}$. The two inputs $\mathbf{z}$ and $c$ are first concatenated into $\mathbb{R}^{33 \times 1}$, and are then transformed by one dense layer with LeakyReLU function into vector $\mathbb{R}^{L d_x \times 1}$, where $L = 15$ and $d_x = 15$. The LeakyReLU function is represented by

$$f(\mathrm{x}) = \begin{cases} x, & \text{if } x \geq 0 \\ \alpha x, & \text{if } x < 0 \end{cases}, \tag{19}$$

where $\alpha$ is the slope coefficient when the value of neuron x is negative. Then, the vector is reshaped into the matrix $\mathbb{R}^{L \times d_m}$ and are linearly transformed into the embedding sequence $\mathbf{X}_{\text{embedding}} \in \mathbb{R}^{L \times d_x}$ with one dense layer. The parameter $d_x = 128$ is the dimension of the embedding representation for the MPC sequence. The embedding sequence is then transformed by the positional encoding, to encode the position information into the sequence $\mathbf{X}$. The operation can be represented by

$$\mathbf{X} = \mathbf{X}_{\text{embedding}} + \mathbf{PE} \tag{20}$$

where $\mathbf{PE} \in \mathbb{R}^{L \times d_x}$ is the learned positional information of the sequence $\mathbf{X}$. Furthermore, the encoded sequence is forwarded to the transformer encoder structure as introduced in "Methods". Following the transformer structure, one Flatten layer and two dense layers are applied to get the output of generator $\hat{\mathbf{x}} \in \mathbb{R}^{60 \times 1}$. The two dense layers have 240 and 60 neurons, respectively. Then, together with the condition variable $c$, the fake channel $\hat{\mathbf{x}}$ or real channel $\mathbf{x} \in \mathbb{R}^{60 \times 1}$ is passed to the discriminator.

The structures of the discriminator and generator are symmetric, with the similar embedding and transformer encoder structure, except that the noise vector in the generator is replaced by the real channel or fake channel in the discriminator. In the Embedding layer, the channel and condition variable are concatenated and transformed. Then, the position encoding learns the position information. Afterwards, the transformer encoder structure is applied. Next, the output of the transformer structure is transformed by two dense layers both with only one neuron. Finally, the discriminator applies the Sigmoid activation function defined as

$$f(x) = \frac{1}{1 + e^{-x}}, \tag{21}$$

which bounds the output of the discriminator between 0 and 1, to represent the probability that the input channel is real.

**Training of T-GAN and TT-GAN**

The training procedure for the proposed T-GAN network is explained in detail as follows. First, the input noise vector for T-GAN is generated using a 32-dimensional multivariate normal distribution, offering flexibility in transforming the noise into the desired distribution. Then, the generator network employs the Stochastic Gradient Descent (SGD) optimizer to ensure its generalization capability, while the discriminator network utilizes the Adaptive Moment Estimation (Adam) optimizer to adaptively control its learning process for fast convergence. The two optimizers are both configured with a learning rate of 0.0001 to ensure stable training. Moreover, the gradient penalty parameter $\lambda$ in equation (6) is set to 10, proving effective to avoid the gradient exploding problem in the training process. The T-GAN is trained using a simulation dataset generated by QuaDRiGa containing 10,000 channels, and the training spans 10,000 epochs. An epoch is defined as a complete pass through the training dataset, with the generator being trained once and the discriminator three times per epoch.

For TT-GAN, the transfer learning process starts by initializing the model with weights from T-GAN, providing a robust starting point for further training. The model is then fine-tuned on real measurement data containing 21 channels for an additional 10,000 epochs, using the same training settings as T-GAN, including the gradient penalty term, optimizers, and learning rate. L2-SP regularization, as described in equation (7), is applied to the parameters to prevent overfitting and ensure the model maintains generalization. This process allows TT-GAN to adapt to new data while retaining valuable features previously learned.

Both T-GAN and TT-GAN are implemented on a Linux server equipped with an AMD Ryzen Threadripper 3990X 64-Core Processor and four NVIDIA GeForce RTX 3090 GPUs, providing the necessary computational power for efficient training.

## Data Availability

All data needed to evaluate the conclusions in the paper are present in the paper and/or the Supplementary Materials. The simulation datasets used in this study are available at https://github.com/huzhengdong/TT-GAN. The measurement data that support the findings of this study are proprietary datasets under collaboration agreements, which can be accessed at https://sites.ji.sjtu.edu.cn/twcd/.

## Code Availability

The full simulation code used for this study can be available at https://github.com/huzhengdong/TT-GAN. The deep learning models are implemented in Python with the framework of TensorFlow.